\documentclass[amsmath, aps,prc,twocolumn,groupedaddress, showkeys, showpacs]{revtex4}
\usepackage{graphicx}

\begin{document}

\title{Entrainment parameters in cold superfluid neutron star core}

\author{Nicolas Chamel}
\affiliation{Copernicus Astronomical Center, Polish Academy of Science, ul. Bartycka 18, 00-716 Warszawa, Poland}
\affiliation{LUTH, Paris Observatory, 5 place Jules Janssen, 92195 Meudon, France}
\author{Pawel Haensel}
\affiliation{Copernicus Astronomical Center, Polish Academy of Science, ul. Bartycka 18, 00-716 Warsaw, Poland}

\date{\today}

\begin{abstract}

Hydrodynamical simulations of neutron star cores, based on a
two fluid description in terms of a neutron-proton superfluid mixture,
require the knowledge of the Andreev-Bashkin entrainment
matrix which relates the momentum  of one constituent to the
currents of both constituents. This matrix is derived for
arbitrary nuclear asymmetry at zero temperature  and in the
limits of small relative currents in the framework of the
energy density functional theory. The Skyrme energy density
functional is considered as a particular case. General
analytic formulae for the entrainment parameters and various
corresponding effective masses are obtained. These formulae
are applied to the liquid core of a neutron star, composed of
an homogeneous plasma of nucleons, electrons and possibly muons
in beta equilibrium.

\end{abstract}


\maketitle

\section{Introduction}

In a standard model of neutron star core,  matter is a uniform
plasma consisting of neutrons, of number density $n_n$, and a
small admixture of protons and electrons, of number densities
$n_p$ and $n_e$, respectively.  Electrons ensure the overall
stability of the star by the condition of electroneutrality,
$n_e=n_p$,  but play a negligible role in the mass
transport because their mass is very small compared to the
nucleon mass. If the electron Fermi energy exceeds the muon rest mass,
muons are present in matter, but their density is always
smaller than that  of electrons. The electroneutrality
condition is then $n_e+n_\mu=n_p$. The electrically charged
particles are strongly coupled to the magnetically braked
solid crust. This builds up a lag between the neutrons and the
protons which is only restored through glitch events. Neutron
star cores are therefore described within two fluid models in
terms of neutron and proton components which are superfluid in
some density range when the temperature of the star falls
below the corresponding critical temperatures.

As a result of the nucleon-nucleon interactions, the momentum
$\pmb{\pi}_q$ of each nucleon  is not simply given by the
corresponding velocity $\pmb{v}_q$ (we use the convention
that $q=n,p$ for neutron, proton respectively) times the mass
$m$ (in the following we shall neglect the small mass
difference between neutrons and protons) but in general is a
linear combination of the neutron and proton velocities. This
is the so called Andreev-Bashkin entrainment effect
\cite{andreev76}. The discussions in the literature have been
usually obscured by the confusion between momentum and
velocity. Traditionally one introduces ``superfluid
velocities'' $\pmb{V}_q$ defined by
\begin{equation}\label{eq.1}
\pmb{V}_q = \pmb{\pi}_q/m
\end{equation}
in which it is recalled that the momenta ${\pmb \pi}_{q}$ are
defined by the partial derivative with respect to the nucleon
current $n_q {\pmb v}_q$ of the Lagrangian density $\Lambda\{
n_q, n_q {\pmb v}_q \}$ of the system
\cite{carterchamel04,prix04}. The mass current of some given
nucleon species $q$
\begin{equation}
\label{eq.2} {\pmb \rho}_{q}=\rho_q {\pmb v}_q \, , \
\rho_q=n_q m
\end{equation}
is then expressible as
\begin{equation}
\label{eq.3} {\pmb \rho}_{q} =\sum_{q^\prime} \rho_{q
q^\prime} {\pmb V}_{q^\prime}
\end{equation}
in which $\rho_{q q^\prime}$ is the (symmetric) entrainment or
mass density matrix. Only one of these matrix elements has to
be specified since the other elements can be obtained from the
identities due to Galilean invariance
\begin{equation}
\rho_{nn}+\rho_{np}=\rho_n \, , \ \rho_{pp}+\rho_{pn}=\rho_p \, .
\end{equation}
This matrix is a necessary ingredient in dynamical simulations of neutron star cores, such as
for instance the study of oscillation modes.
The (static) equation of state and the entrainment matrix are usually
obtained using different microscopic models. In earlier
calculations and even recently, the mass density matrix is
postulated to have some density dependence whose parameters
are determined from rough estimates.

Comer \textit{et al.} \cite{comer03} have built a
self-consistent equation of state in the framework of a
minimal relativistic $\sigma-\omega$ mean field model,
ignoring non linear couplings between the meson fields which
are however essential in order to reproduce nuclear properties such as
the incompressibility of nuclear matter. They have obtained
semi analytical formulae for the entrainment parameters in the
limit of small fluid velocities (compared to that of light),
which even within this simple mean field model take a rather
complicated form. It is not clear that analytical formulae
could still be obtained with realistic relativistic mean field
models, taking into account self meson couplings and including
as well the $\rho$ meson which is required for a correct
treatment of the symmetry energy.

Despite the fact that non relativistic mean field models have
been widely applied in the study of terrestrial nuclei and in
neutron stars, there has been no attempt so far to apply these
models for the calculation of the mass density matrix. The
purpose of the present work is therefore to fill this gap and
to further investigate the density dependence of the various
entrainment parameters and effective masses that have been
introduced in the literature.

\section{Entrainment in a mixture of Fermi liquids}
At zero temperature the entrainment effects have been shown to
be independent of the nucleon pairing, giving rise to
superfluidity \cite{CCH05, Gusakov05}. Even at finite
temperatures well below the critical temperatures for the
onset of superfluidity, pairing as well as thermal effects are
very small \cite{Gusakov05}. We can therefore ignore pairing interactions and
restrict ourself to the limit of zero temperature.

Borumand \textit{et al.} \cite{borumand96} have shown how to
obtain the entrainment matrix of a neutron-proton mixture in
the framework of the Landau Fermi liquid theory. In what
follows, we will limit ourselves to spin-unpolarized nuclear
matter. Therefore, spin indices will not appear in our
formulae, and all quantities are to be understood as
spin-averages.
Under our assumptions, the change in the total energy density of the
system due to a small current is expressed as
\begin{multline}
\label{eq.3b}
{\cal E} = 2\sum_{q} \int\frac{{\rm d}^3 \pmb k}{(2\pi)^3} e^{(q)} \{ {\pmb k} \}
\delta \widetilde{n}^{(q)}\{ {\pmb k} \}
\\+ 2\sum_{q, q^\prime} \int \frac{{\rm d}^3 {\pmb k}}{(2\pi)^3} \int \frac{{\rm d}^3 {\pmb k}^\prime}{(2\pi)^3}
f^{q q^\prime}\{ {\pmb k}, {\pmb k^\prime} \}
\delta \widetilde{n}^{(q)}\{ {\pmb k} \}
\delta \widetilde{n}^{(q^\prime)}\{ {\pmb k^\prime} \} \, ,
\end{multline}
in which $e^{(q)} \{ {\bf k} \}$ is the energy of a
quasiparticle ($q=n,p$ for neutron and proton respectively) of
wave vector $\pmb k$, $ f^{q q^\prime}\{ {\pmb k}, {\pmb
k^\prime} \}$ is the (spin averaged) interaction between the
quasiparticles. Moreover, $\delta\widetilde{n}^{(q^\prime)}\{
{\pmb k^\prime} \}$ denotes the change in the distribution
function of quasiparticle states from that of the static (zero
current)  ground state characterized by the Heaviside
functions $\Theta\{k^{(q)}_{\rm F} - k\}$, where $k^{(q)}_{\rm F}$
is the Fermi momentum  (in the units of $\hbar$) $k_{\rm
F}^{(q)}=(3 \pi^2 n_q)^{1/3}$.  In the presence of neutron and
proton currents, the corresponding Fermi surfaces are
displaced by a vector ${\pmb Q}_q$. In the limit of small
currents $Q_q \ll k_{\rm F}^{(q)}$, writing the ``superfluid
velocities'' from (\ref{eq.1}) as
\begin{equation}\label{eq.4}
{\pmb V}_q = \hbar {\pmb Q}_q/m
\end{equation}
it can be shown that the mass current ${\pmb \rho}_q=\rho_q
{\pmb v_q}$ of each nucleon species is linearly related to
both the neutron and proton superfluid velocities
\begin{equation}
\label{eq.5}
{\pmb \rho}_q= \sum_{q^\prime}
\rho_{qq^\prime}{\pmb V}_{q^\prime} \, ,
\end{equation}
where the (symmetric) entrainment matrix $\rho_{q q^\prime}$
is given by

\begin{equation}
\label{ABmatrix}
\rho_{q q^\prime}=\sqrt{\rho_q \rho_{q^\prime}}
\frac{m}{\sqrt{m_q^\oplus m_{q^\prime}^\oplus}} (\delta_{q
q^\prime}+{\cal F}_1^{q q^\prime}/3) \, .
\end{equation}
The (Landau) effective mass $m_q^\oplus$ and the dimensionless
Landau parameters ${\cal F}_\ell^{q q^\prime}$ are defined
respectively by
\begin{equation}
\label{eq.9}
\frac{1}{m_q^\oplus}=\frac{1}{\hbar^2 k_{_{\rm F}}^{_{(q)}}}
\frac{{\rm d}e}{{\rm d} k}\biggr\vert_{k=k_{\rm F}^{(q)}}\, ,
\end{equation}
\begin{equation}
\label{eq.11}
{\cal F} _\ell^{qq^\prime} =
\sqrt{{\cal N}_q {\cal N}_{q^\prime}} f^{qq^\prime}_\ell \, ,
\end{equation}
in which ${\cal N}_q$ is the density of
quasiparticle states at the Fermi surface,
\begin{equation}
\label{eq.12}
{\cal N}_q=\frac{m_q^\oplus k_{\rm F}^{(q)}}{\hbar^2 \pi^2}
\,,
\end{equation}
and the parameters $f_\ell^{q q^\prime}$ are obtained from the
Legendre expansion of the spin averaged quasiparticle
interaction
\begin{equation}
\label{eq.10}
f^{q q^\prime}\{ {\pmb  k}, {\pmb k^\prime} \} =
\sum_{\ell} f^{qq^\prime}_\ell P_\ell\{ \cos
\theta \}
\end{equation}
where $\theta$ is the angle between the wave vectors $\pmb k$
and $\pmb k^\prime$ lying on the corresponding Fermi surface.

Alternative formulae of the entrainment matrix \cite{alpar84}
have been used in the literature, based on the decomposition
of the Landau effective masses in the form
\begin{equation}
m_n^\oplus = m+\delta m_{nn}^\oplus + \delta m_{np}^\oplus \, ,
\end{equation}
\begin{equation}
m_p^\oplus = m+\delta m_{pp}^\oplus + \delta m_{pn}^\oplus
\end{equation}
where the various contributions to the effective masses are
related to the Landau parameters by the simple formula
\cite{sjoberg76}
\begin{equation}
\delta m_{q q^\prime}^\oplus = \frac{1}{3} {\cal F}_1^{q
q^\prime} m \sqrt{\frac{n_{q^\prime} m_q^\oplus}{n_q
m_{q^\prime}^\oplus}} \, .
\end{equation}

The mass density matrix can then be equivalently written
explicitly as
\begin{equation}
\rho_{nn}=\rho_n \frac{m+\delta m^\oplus_{nn}}{m_n^\oplus}
\end{equation}
\begin{equation}
\rho_{pp}=\rho_p \frac{m+\delta m^\oplus_{pp}}{m_p^\oplus}
\end{equation}
\begin{equation}
\rho_{np}=\rho_{pn}=
\rho_n\frac{\delta m^\oplus_{np}}{m_n^\oplus}=
\rho_p\frac{\delta m^\oplus_{pn}}{m_p^\oplus} \, .
\end{equation}
It should be remarked that in the formulae provided by Sauls
(see \cite{alpar84}), the terms proportional to $\delta
m^\oplus_{nn}$ and $\delta m^\oplus_{pp}$ in the expressions
for $\rho_{nn}$ and $\rho_{pp}$ are omitted and therefore
those formulae violate Galilean invariance.

The quasiparticle energies $e^{(q)} \{ {\pmb k} \} $ and the
quasiparticle  interaction $f^{q q^\prime}\{ {\pmb k}, {\pmb
k^\prime} \}$ can be deduced from a microscopic approach.
The solution of the many body problem, starting from the bare
nucleon-nucleon interactions is very difficult. We shall here
adopt a simpler approach based on self-consistent mean field
models with phenomenological effective interactions (for a
review, see for instance \cite{bender03}), which have been
very successful in describing the nuclear properties of
terrestrial nuclei. Such mean field models have also been
widely applied in the context of neutron stars.
\section{Landau parameters in the energy density functional theory}
We shall calculate in this section the Landau parameters for
asymmetric nuclear matter in the framework of the
Hohenberg-Kohn-Sham energy density functional theory\cite{hohenbergkohn64, kohnsham65}.

The energy density functional
for spin-unpolarized homogeneous nuclear matter is written as
a sum of the isoscalar ($T=0$) and isovector ($T=1$) terms
\cite{bender03}
\begin{equation}\label{eq.15}
{\cal E} =
\sum_{T=0,1} \delta_{_{T0}} \frac{\hbar^2}{2 m}
 \tau_{_T} + C_{_T}^n\{ n_{\rm b}\} n_{_T}^2 + C_{_T}^\tau n_{_T}
 \tau_{_T} + C_{_T}^j {{\pmb j}_{_T}}^2 \, .
\end{equation}
The isoscalar and isovector parts of some quantity for a
nucleon system is given respectively by the sum and the
difference between the neutron and proton contributions. For
example, the isoscalar and isovector densities are given by
$n_0=n_n + n_p=n_{\rm b}$ and $n_1=n_n-n_p$ respectively.

The nucleon density $n_q$, kinetic energy density $\tau_q$ (in the
units of $\hbar^2/2m$), and nucleon current ${\pmb
j}_q$ are expressible in terms of the nucleon distribution
function $\widetilde{n}^{(q)}\{ {\pmb k} \}$ by
\begin{equation}\label{eq.16}
n_q =  \int \frac{{\rm d}^3\pmb k}{(2\pi)^3}  \widetilde{n}^{(q)}\{ {\pmb k} \}
\end{equation}
\begin{equation}\label{eq.17}
\tau_q =  \int \frac{{\rm d}^3\pmb k}{(2\pi)^3}  k^2 \widetilde{n}^{(q)}\{ {\pmb k} \}
\end{equation}
\begin{equation}\label{eq.18}
{\pmb j}_q = \int \frac{{\rm d}^3\pmb k}{(2\pi)^3}  {\pmb k}\, \widetilde{n}^{(q)} \{
{\pmb k} \} \, .
\end{equation}

Energy density functionals of the form (\ref{eq.15}) can be
obtained in the Hartree-Fock approximation with effective
contact nucleon-nucleon interactions $\hat v \{ {\pmb r}_1,
{\pmb r}_2 \}$ of the Skyrme type, whose standard
parametrisations (ignoring spin-orbit terms which are
irrelevant in the present case) are
\begin{multline}\label{eq.13b}
\hat v \{ {\pmb r}_1, {\pmb r}_2 \}= t_0(1+x_0 \hat P_\sigma)
\delta\{ {\pmb r}_1 - {\pmb r}_2 \} \\ + \frac{1}{2} t_1
(1+x_1 \hat P_\sigma)\bigl(  \hat{\pmb k}^{\dagger2} \delta\{
{\pmb r}_1 - {\pmb r}_2 \} +\delta\{ {\pmb r}_1 - {\pmb r}_2 \}
\hat{\pmb k}^2 \bigr) \\ +t_2(1+x_2 \hat P_\sigma)
\hat{\pmb k}^{\dagger} \cdot \delta\{ {\pmb r}_1 -
{\pmb  r}_2 \} \hat{\pmb k} \\ +\frac{1}{6} t_3
(1+x_3 \hat P_\sigma) \delta\{ {\pmb r}_1
 - {\pmb  r}_2 \} n_{\rm b}\bigl\{
 \frac{ {\pmb r}_1+{\pmb  r}_2}{2}
\bigr\}^\gamma \, ,
 \end{multline}
where $\hat P_\sigma=(1+{\pmb \sigma}_1 \cdot {\pmb \sigma}_2
)/2$ is the spin exchange operator and $\hat{\pmb k}=-{\rm i}
(\nabla_1 - \nabla_2)/2$. The density dependent term
proportional to $t_3$ represents the effects of three body
interactions. The coefficients $C_{_T}^n\{n_b\}$, $C_{_T}^\tau$ and
$C_{_T}^j$ can then be expressed in terms of the parameters of
the Skyrme interaction (see Appendix). It should be stressed
however that the functional (\ref{eq.15}) is more general than
the Skyrme functional. In particular, the coefficients
$C_{_T}^n\{ n_b\}$ can be any function of the baryon density
$n_{\rm b}$.

The single particle energies are obtained from the functional derivative of
the energy density
\begin{equation}\label{eq.19}
e^{(q)} \{ {\pmb k} \} = \frac{\delta \cal E}{\delta
\widetilde{n}^{(q)} \{ {\pmb  k} \}} \biggr \vert_0
\end{equation}
where the zero subscript indicates that the functional
derivative is evaluated  in  the static ground state (in which
the currents ${\pmb j}_q$ vanish),  characterized by the
distribution function
\begin{equation}\label{eq.28}
\widetilde{n}_0^{(q)} \{ {\pmb k} \}=
 \Theta\{ k_{\rm F}^{(q)} - k \} \, .
\end{equation}

Substituting the functional (\ref{eq.15}) yields
\begin{equation}\label{eq.20}
e_q\{ k \} = \frac{\hbar^2 k^2}{2 m_q^\oplus} + U_q \, ,
\end{equation}
in which the effective mass $m_q^\oplus$ (using the same
symbol as for the Landau effective mass defined by
(\ref{eq.9}) since both definitions coincide) and the
single-quasiparticle potential  $U_q$ are given by
\begin{equation}
\label{eq.21}
\frac{\hbar^2}{2 m_q^\oplus} =
\frac{\delta {\cal E} }{\delta \tau_q} =
\frac{\hbar^2}{2 m}+(C_0^\tau-C_1^\tau) n_{\rm b}
 +2 C_1^\tau n_q
\end{equation}
\begin{multline}
\label{eq.22}
U_q = \frac{\delta {\cal E} }{\delta n_q}
=4 C_1^n n_q - 2 n_b(C_0^n-C_1^n)\\
 + (C_0^\tau -C_1^\tau)\tau_{\rm b} +2 C_1^\tau \tau_q
 + \frac{{\rm d} C_0^n}{ {\rm d} n_{\rm b}} n_{\rm b}^2
+ \frac{{\rm d} C_1^n}{ {\rm d}  n_{\rm b}}(2 n_q - n_{\rm b})^2 \,
,
\end{multline}
where $\tau_{\rm b}=\tau_n+\tau_p$.

The quasiparticle interaction, calculated as a second
functional derivative of the energy functional (\ref{eq.15}),
is
\begin{equation}
\label{eq.23}
f^{q q^\prime}\{ {\pmb k}, {\pmb  k^\prime} \}= \frac{\delta^2
\cal E}{\delta \widetilde{n}^{(q)}\{ {\pmb  k} \}\delta
\widetilde{n}^{(q^\prime)}\{ {\pmb k}^\prime \}} \biggr
\vert_0 \, ,
\end{equation}
and contains only $\ell=0$ and $\ell=1$ components.

The non-vanishing Landau parameters are found to be
expressible as
\begin{eqnarray}
f_0^{nn} =2 k_{\rm F}^{(n)2}(C_0^\tau+C_1^\tau)
+2(C_1^n+C_0^n)+\frac{{\rm d}C_0^n}{{\rm d}n_b}
4 n_{\rm b}\nonumber
\\
+4\frac{{\rm d}C_1^n}{{\rm d}n_{\rm b}}
(n_n-n_p)+\frac{{\rm d}^2C_0^n}{{\rm d}n_{\rm b}^2}n_{\rm b}^2
+\frac{{\rm d}^2 C_1^n}{{\rm d} n_{\rm b}^2}(n_n-n_p)^2\, ,
\end{eqnarray}
\begin{eqnarray}
f_0^{pp} =2 k_{\rm F}^{(p)2}(C_0^\tau+C_1^\tau)
+2(C_1^n+C_0^n)+\frac{{\rm d}C_0^n}{{\rm d}n_{\rm b}}
4 n_{\rm b}\nonumber\\
-4\frac{dC_1^n}{{\rm d}n_{\rm b}}(n_n-n_p)
+\frac{{\rm d}^2C_0^n}{{\rm d}n_{\rm b}^2}n_{\rm b}^2
+\frac{{\rm d}^2 C_1^n}{{\rm d} n_{\rm b}^2}(n_n-n_p)^2\, ,
\end{eqnarray}
\begin{widetext}
\begin{equation}
f_0^{np} = f_0^{pn}=(k_{\rm F}^{(n)2}+ k_{\rm F}^{(p)2})
(C_0^\tau-C_1^\tau)+2(C_0^n-C_1^n)
+\frac{{\rm d}C_0^n}{{\rm d}n_b}4 n_{\rm b}
+\frac{{\rm d}^2C_0^n}{{\rm d}n_{\rm b}^2}n_{\rm b}^2
+\frac{{\rm d}^2 C_1^n}{{\rm d} n_{\rm b}^2}(n_n-n_p)^2{}
\end{equation}
\end{widetext}
\begin{equation}
f_1^{nn} = 2(C_0^j +C_1^j) k_{\rm F}^{(n)2} \, ,
\end{equation}
\begin{equation}
f_1^{pp} = 2(C_0^j +C_1^j) k_{\rm F}^{(p)2} \, ,
\end{equation}
\begin{equation}
f_1^{np} =f_1^{pn}=2(C_0^j-C_1^j) k_{\rm F}^{(n)} k_{\rm F}^{(p)} \, .
\end{equation}

These formulae agree with those of Bender \textit{et al.}
\cite{bender02} for the limiting case of symmetric nuclear
matter (using standard notations $f_\ell^{nn}
=f_\ell^{pp}=f_\ell+f_\ell^\prime$ and $f_\ell^{np} = f_\ell -
f_\ell^\prime$) and generalize the results of Blaizot \&
Haensel \cite{blaizothaensel81} for asymmetric nuclear matter
to any energy density functional of the form (\ref{eq.15}).

It should be remarked in particular that for any such
functional (\ref{eq.15}), the parameters $f_1^{nn}$ and
$f_1^{pp}$ are related by
\begin{equation}
\frac{f_1^{nn}}{f_1^{pp}}=\biggl(\frac{n_n}{n_p}\biggr)^{2/3} \, .
\end{equation}

The corresponding dimensionless $\ell=1$ Landau parameters
can be expressed in compact form as
\begin{equation}
\label{eq.27}
{\cal F}_1^{qq^\prime} = 3 \widetilde{\alpha}_{qq^\prime}
\sqrt{n_q m_q^\oplus n_{q^\prime} m_{q^\prime}^\oplus} \, ,
\end{equation}
in which the coefficients $\widetilde\alpha_{q q^\prime}$
are defined by
\begin{equation}
\label{eq.25}
\widetilde\alpha_{nn} = \widetilde\alpha_{pp}
=\frac {2}{\hbar^2}(C_0^j +C_1^j )
\end{equation}
\begin{equation}
\widetilde\alpha_{np}=\widetilde\alpha_{pn}
=\frac{2}{\hbar^2}(C_0^j-C_1^j) \, .
\end{equation}

We conclude this section by remarking that in a general case
of asymmetric nuclear matter (i.e., with $n_n\neq n_p$) the
$\ell=1$ Landau parameters can be uniquely determined in terms
solely of the effective masses as
\begin{equation}
{\cal F}_1^{np}=\frac{3}{m} \frac{\sqrt{n_n m_n^\oplus n_p m_p^\oplus}}
{n_p^2 -n_n^2} \left[n_p\left(1-\frac{m}{m_n^\oplus}\right)-n_n\left(1-\frac{m}{m_p^\oplus}\right)\right]
\end{equation}
\begin{equation}
{\cal F}_1^{nn}=3\frac{n_n}{n_p^2-n_n^2} \frac{m_n^\oplus}{m}\left[ n_p\left(1-\frac{m}{m_p^\oplus}\right)-n_n\left(1-\frac{m}{m_n^\oplus}\right)\right]
\end{equation}
\begin{equation}
{\cal F}_1^{pp}=3\frac{n_p}{n_p^2-n_n^2} \frac{m_p^\oplus}{m}\left[ n_p\left(1-\frac{m}{m_p^\oplus}\right)-n_n\left(1-\frac{m}{m_n^\oplus}\right)\right]
\end{equation}

\section{Entrainment matrix and effective masses}
Subtituting the expressions (\ref{eq.27}) of the Landau parameters obtained
in the previous section, the entrainment matrix elements (\ref{ABmatrix}) can
be seen to be expressible as
\begin{equation}
\label{eq.30}
\rho_{q q^\prime} = \rho_q \frac{m}{m_q^\oplus}
\delta_{q q^\prime} + \widetilde{\alpha}_{q q^\prime}
\rho_q \rho_{q^\prime} \, ,
\end{equation}
or more explicitly in terms of the mass densities $\rho_n$ and
$\rho_p$
\begin{equation}\label{eq.31}
\rho_{nn} = \rho_n ( 1-\widetilde{\alpha}_{np} \rho_p)
\end{equation}
\begin{equation}\label{eq.32}
\rho_{pp} =  \rho_p ( 1-\widetilde{\alpha}_{np} \rho_n)
\end{equation}
\begin{equation}\label{eq.33}
\rho_{np}=\widetilde{\alpha}_{np} \rho_n\rho_p \, .
\end{equation}
It is readily seen that the formulae (\ref{eq.31}-\ref{eq.33})
imply  basic property of the entrainment matrix, namely
$\rho_{nn}+\rho_{np}=\rho_n$ and $\rho_{pp}+\rho_{np}=\rho_p$,
which guarantees the Galilean invariance of the two fluid
model.

Two other kinds of effective masses, different from
 the Landau
effective masses $m^\oplus_q$,  have been introduced in the literature.
Effective nucleon masses can be defined from the mass density matrix
elements by setting
\begin{equation}\label{eq.34}
\frac{\rho_{qq}}{\rho_q} = \frac{m}{m^q_\sharp} \, ,
\end{equation}
in such a way that in the proton \emph{momentum} rest frame (${\pmb
V_p}=0$) we have ${\pmb \pi}_n=m_\sharp^n {\pmb v}_n$ and similarly
in the neutron \emph{momentum} rest frame,  ${\pmb \pi}_p=m_\sharp^p {\pmb v}_p$.

These effective masses have a very simple density dependence
as shown on the formulae
\begin{equation}\label{eq.35}
\frac{m_\sharp^n}{m} = \frac{1}{1-\widetilde{\alpha}_{np} \rho_p}
\end{equation}
\begin{equation}\label{eq.36}
\frac{m_\sharp^p}{m} =\frac{1}{ 1-\widetilde{\alpha}_{np} \rho_n} \, .
\end{equation}
The $\sharp$-effective masses differ from the Landau
quasiparticle effective masses, and are related to the latter
ones by
\begin{equation}\label{eq.37}
\frac{m}{m_\sharp^q} = \frac{m}{m_q^\oplus}
+\widetilde{\alpha}_{qq} \rho_q \, ,
\end{equation}
due to the non-vanishing quasiparticle interactions. The extra
term on the right hand side can be interpreted as resulting from  the
backflow induced by the motion of the quasiparticles.

Alternatively one can introduce effective masses $m^q_\star$
such that in the proton rest frame (meaning ${\pmb v}_p=0$) we
have ${\pmb \pi}_n=m^n_\star {\pmb v}_n$ and similarly in the neutron rest frame
${\pmb \pi}_p=m^p_\star {\pmb v}_p$. These
effective masses are given by
\begin{equation}\label{eq.38}
\frac{m^n_\star}{m}=
\frac{1-\widetilde{\alpha}_{np}\rho_n}
{1-\widetilde{\alpha}_{np}\rho_{\rm b}}\, ,
\end{equation}
\begin{equation}\label{eq.39}
\frac{m^p_\star}{m}=\frac{1-
\widetilde{\alpha}_{np}\rho_p}
{1-\widetilde{\alpha}_{np}\rho_{\rm b}} \, .
\end{equation}
where $\rho_{\rm b}=\rho_n+\rho_p$.
The effective masses of the different kinds are related by
\begin{equation}\label{eq.40}
m_\sharp^n-m = (m_\star^n - m)\left[1+\frac{n_n}{n_p}
\left( \frac{m_\star^n}{m}-1\right) \right]^{-1} \, ,
\end{equation}
\begin{equation}\label{eq.41}
m_\sharp^p-m = (m_\star^p - m)\left[1+\frac{n_p}{n_n}
\left( \frac{m_\star^p}{m}-1\right) \right]^{-1} \, .
\end{equation}

These formulae show that, as pointed out by Prix \textit{et al.}
 \cite{prix02}, in the limit of very small proton fraction
 $n_p/n_{\rm b} \ll 1$, as relevant in the liquid core of neutron stars,
 we shall have $m_\star^n \sim m_\sharp^n \sim m$ and $m_\star^p \sim
m_\sharp^p$. We shall compute more accurate values of the effective masses
in section \ref{ns}.

In studies of neutron star cores, the non diagonal
entrainment matrix elements $\rho_{np}$ has often been parametrised as
\begin{equation}
\label{eq.42}
\rho_{np}=-\epsilon \rho_n
\end{equation}
in which the dimensionless parameter $\epsilon$ was taken as a
constant \cite{lindblom00, lee03, yoshida03}. Other authors
\cite{prixrieutord02, prix02, prixcomer04} have suggested instead
to set the dimensionless parameters defined by
\begin{equation}
\label{eq.43}
\varepsilon_q=1-\frac{m^q_\star}{m} \, ,
\end{equation}
as constants. However comparison with (\ref{eq.33}), (\ref{eq.38})
and (\ref{eq.39}) shows that neither ansatz is satysfying, since
these parameters are found to vary with the densities according to
\begin{equation}
\label{eq.44}
\epsilon=-\widetilde{\alpha}_{np} \rho_p \, .
\end{equation}
\begin{equation}
\label{eq.45} \varepsilon_n=-\frac{\widetilde{\alpha}_{np} \rho_p}
{1-\widetilde{\alpha}_{np}\rho_{\rm b}}
\end{equation}
\begin{equation}
\label{eq.46} \varepsilon_p=-\frac{\widetilde{\alpha}_{np} \rho_n}
{1-\widetilde{\alpha}_{np}\rho_{\rm b}} \, .
\end{equation}

The effective masses and entrainment parameters seem to diverge at
some points of the $\rho_n-\rho_p$ plane.  However as will be
shown in the next section, once stability constraints are imposed,
these apparent singularities disappear.

\section{Stability of the static ground state and constraints on the entrainment
parameters}
Since the momentum of each nucleon is a linear combination of both
the neutron  and proton currents, this means that the
corresponding dynamical contribution to the Lagrangian density of
the system $\Lambda_{\rm dyn}={\cal E}_{\rm dyn}$ is a bilinear
symmetric form of the currents. It is therefore readily seen that
this dynamical contribution is expressible in terms of the
Andreev-Bashkin entrainment matrix elements as
\begin{equation}\label{eq.47}
{\cal E}_{\rm dyn} =\frac{1}{2} (\rho_{nn} {\pmb V}_n^2 + 2
\rho_{np} {\pmb V}_n \cdot {\pmb V}_p + \rho_{pp} {\pmb V}_p^2 )
\, .
\end{equation}
As a result the total energy density ${\cal E}$ of the fluid
mixture can be written as the sum of the dynamical contribution,
${\cal E}_{\rm dyn}$, and an internal static contribution ${\cal
E}_{\rm ins}$ which only depends on the densities: ${\cal E}={\cal
E}_{\rm dyn}+{\cal E}_{\rm ins}$ .

The static ground state of the system is stable  if the term
${\cal E}_{\rm dyn}$ is strictly positive. This means that the
entrainment matrix must be positive definite (the minimum energy
state thus being obtained by the vanishing of the superfluid
velocities or equivalently of the currents, \textit{i.e.} ${\cal
E}_{\rm dyn}=0$), which means that its eigenvalues must be
strictly positive. This condition entails that the matrix elements
(\ref{eq.31}-\ref{eq.33}) should obey
\begin{equation}\label{eq.49}
\rho_{nn}+\rho_{pp} >0\, ,
\end{equation}
\begin{equation}\label{eq.50}
\rho_{np}^2 < \rho_{nn} \rho_{pp} \, .
\end{equation}

These conditions lead to contraints on the $\ell=1$ Landau parameters
(using the other constraint that the Landau effective masses $m_n^\oplus$
and $m_p^\oplus$ have to be positive)
\begin{equation}
{\cal F}^{nn}_1>-3\, , \hskip 0.3cm {\cal F}^{pp}_1>-3
\end{equation}
together with
\begin{equation}
\left(1+\frac{1}{3}{\cal F}^{nn}_1\right)\left(1+\frac{1}{3}{\cal F}^{pp}_1\right) > \left(\frac{{\cal F}_1^{np}}{3}\right)^2 \, .
\end{equation}

The stability conditions can also be expressed in terms of effective masses
 \cite{CCH}
\begin{equation}
\label{eq.51} \frac{m^q_\star}{m} > \frac{n_q}{n_{\rm b}} \, ,
\end{equation}
or equivalently
\begin{equation}
\label{eq.52}
\frac{m^q_\sharp}{m} < \frac{n_b}{n_q} \, .
\end{equation}
In terms of the dimensionless entrainment  parameters
$\varepsilon_q$, these conditions can be expressed as
\begin{equation}
\label{eq.53} \varepsilon_q < 1-\frac{n_q}{n_{\rm b}} \, .
\end{equation}
It should be remarked that  the previous inequalities
are very general and have to be satisfied in \emph{any}
 two fluid models. In the present case,
these conditions also impose a constraint on the energy functional
(\ref{eq.15}) from which the entrainment matrix is derived. Since
the conditions (\ref{eq.49}) and (\ref{eq.50}) must be satisfied
for any neutron and proton densities, this leads to the following
requirement
\begin{equation}
\label{eq.54}
C_0^j\leq C_1^j \, .
\end{equation}
In the particular case of Skyrme functionals, this last condition reads
\begin{equation}\label{eq.55}
t_1 (2+x_1)+t_2 (2+x_2) \geq 0 \, .
\end{equation}

Whenever this condition is fulfilled, it can be seen from
 equations
(\ref{eq.35}-\ref{eq.39}) that for any
neutron and proton densities, the effective masses $m^q_\star$ and
$m^q_\sharp$ are therefore positive and smaller than the bare
nucleon mass
\begin{equation}\label{eq.56}
0< m^q_\star, m^q_\sharp\leq m \, .
\end{equation}

 Combining the latter inequality with (\ref{eq.51}) shows in particular that
\begin{equation}\label{eq.57}
n_q/n_{\rm b}< m^q_\star/m \leq 1 \, .
\end{equation}
Besides,  since the proton fraction is very small inside neutron
stars, it can be seen from (\ref{eq.54}) and the definitions
 (\ref{eq.35}), (\ref{eq.36}), (\ref{eq.38}) and
(\ref{eq.39}) that in this case the neutron effective masses
are always larger than the proton ones
\begin{equation}
m^n_\star > m^p_\star \, , \ \ m^n_\sharp>m^p_\sharp \, .
\end{equation}
Likewise, it can be shown that $\varepsilon_q\geq 0$ which, in association with (\ref{eq.53}), yields
\begin{equation}\label{eq.58}
0\leq \varepsilon_q< 1- \frac{n_q}{n_{\rm b}} \, .
\end{equation}

It is thus found that the entrainment parameters are well behaving
functions of nucleon densities.

\section{Application to neutron star matter in beta equilibrium}
\label{ns}
In the previous section we  have obtained general formulae for
the entrainment parameters and the associated effective masses
for nuclear matter with arbitrary asymmetry. In the present
section we shall apply these formulae to construct a model of
neutron star core. We shall use the SLy4 Skyrme force which
has been specifically devised for astrophysical purposes
\cite{chabanat95, chabanat97, chabanat98, chabanat98err}. In
the framework of a compressible liquid drop model based on the
SLy4 Skyrme energy functional, Douchin \& Haensel
\cite{douchinhaensel01} found that the bottom edge of the
crust corresponds to the baryon density $n_{\rm edge}\simeq
0.076$ fm$^{-3}$. In the following we shall consider the density
domain $n_{\rm edge}<n_{\rm b}<3 n_{\rm s}$, where
$n_{\rm s}=0.16$ fm$^{-3}$ is the nuclear saturation density.

We assume that the
liquid core is composed of an homogeneous plasma of neutrons,
protons and electrons (and muons for baryon densities beyond
some critical threshold) in beta equilibrium
\begin{equation}
n \leftrightarrow p^+ +e^- +\nu_e \, , \ \mu^- \leftrightarrow
e^-+\nu_\mu+ \bar \nu_e \, .
\end{equation}
This means that the chemical potentials of the various species are
related by (assuming that neutrinos escaped from the star)
\begin{equation}
\mu_n=\mu_p+\mu_e \, , \ \mu_e=\mu_\mu~.
\end{equation}
In the Hartree-Fock approximation, chemical potentials of nucleons
are equal to  the corresponding Fermi energies ($q=n,p$)
 including rest mass
energy
\begin{equation}\label{muq}
\mu_q= m_q c^2 + \frac{\hbar^2 k^{(q)2}_F}{2 m_q^\oplus} + U_q \, .
\end{equation}
Considering the leptons as ideal relativistic
Fermi gases,
 the lepton chemical potentials are given by  ($l=e,\mu$)
\begin{equation}\label{mul}
\mu_l= \sqrt{m_l c^2 +\hbar^2 c^2 (3\pi^2 n_l)^{2/3}}
 \, .
\end{equation}
Charge neutrality requires that
\begin{equation}
n_p=n_e+n_\mu \, .
\end{equation}

In equations (\ref{muq})  and (\ref{mul}), we have neglected the
deviations in the chemical potentials due to the existence of non
vanishing currents since the relative velocities are typically very small
compared to the velocities of the various constituents.
For completeness, let us mention that the internal energy density
of the nucleons can be decomposed in the form
\begin{equation}
{\cal E}_{\rm int} = {\cal E}_0 + {\cal E}_{\rm ent}
\end{equation}
in which ${\cal E}_0$ is the functional (\ref{eq.15}) evaluated
in the static ground state with the distribution function
(\ref{eq.28}) including the rest mass energies,
\begin{widetext}
\begin{equation}
{\cal E}_0 \{n_n,n_p\}= n_b m c^2+ \left( \frac{\hbar^2}{2
m}+C_0^\tau n_{\rm b} \right) \frac{3}{5}(3 \pi^2)^{2/3}
(n_n^{5/3}+n_p^{5/3})\\ +C_0^n n_{\rm b}^2 +
C_1^n(n_n-n_p)^2+C_1^\tau(n_n-n_p) \frac{3}{5}(3 \pi^2)^{2/3}
(n_n^{5/3}-n_p^{5/3}) \, , {}
\end{equation}
\end{widetext}
and ${\cal E}_{\rm ent}$  is the entrainment contribution which is
expressible as
\begin{equation}
{\cal E}_{\rm ent} = -\frac{1}{2} \rho_n \varepsilon_n (\delta v)^2
\end{equation}
where $\pmb \delta \pmb v$  is the velocity
difference between neutrons and protons. The entrainment term
is negligibly small compared to the static term ${\cal E}_{\rm
ent}\ll {\cal E}_0$ even for the fastest pulsars and can therefore be neglected.

\begin{figure}
\includegraphics[height=6cm,draft=false]{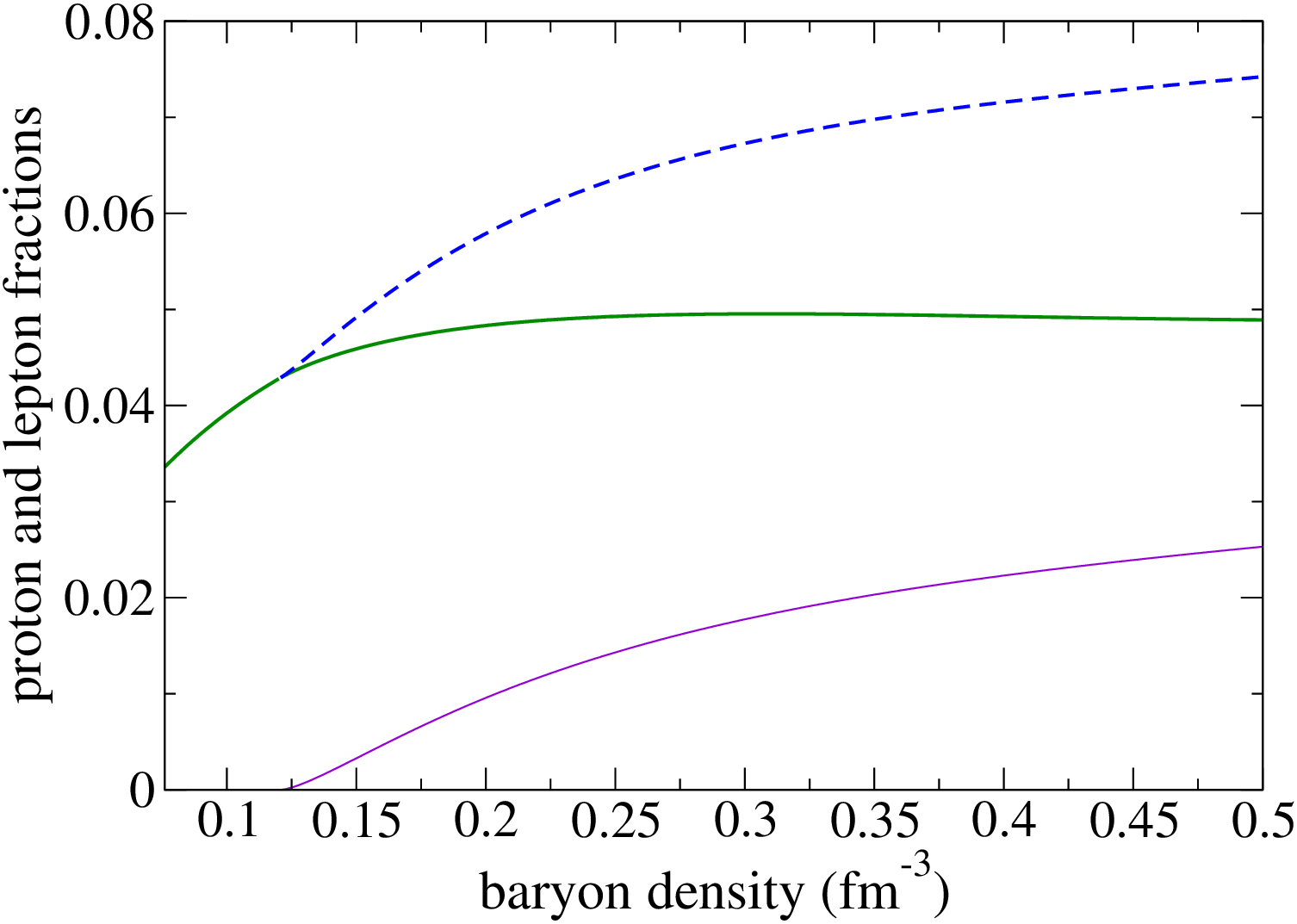}
\caption{\label{fig.composition}(Color online) Equilibrium fractions of protons
(dashed line), electrons (thick line) and muons (thin line) in
neutron star liquid core as a function of the baryon density
$n_b=n_p+n_n$ from the bottom edge of the crust $n_{\rm
edge}\simeq 0.076$ fm$^{-3}$ down to $3 n_{\rm s}$ where $n_{\rm s}=0.16$
fm$^{-3}$ is the nuclear saturation density. The results have been
obtained with the Skyrme SLy4 energy density functional.}
\vskip 0.7cm
\end{figure}

The muons are present in matter when the  electron chemical
potential $\mu_e$ exceeds the muon mass $m_\mu c^2\simeq 105$ MeV.
This occurs at a baryon density $n_{\rm b} \simeq 0.12$ fm$^{-3}$.
In equilibrium, the composition of the  liquid core is therefore
completely determined by the baryon density $n_{\rm b}$ and is
shown on figure \ref{fig.composition}.

\begin{figure}
\includegraphics[height=6cm,draft=false]{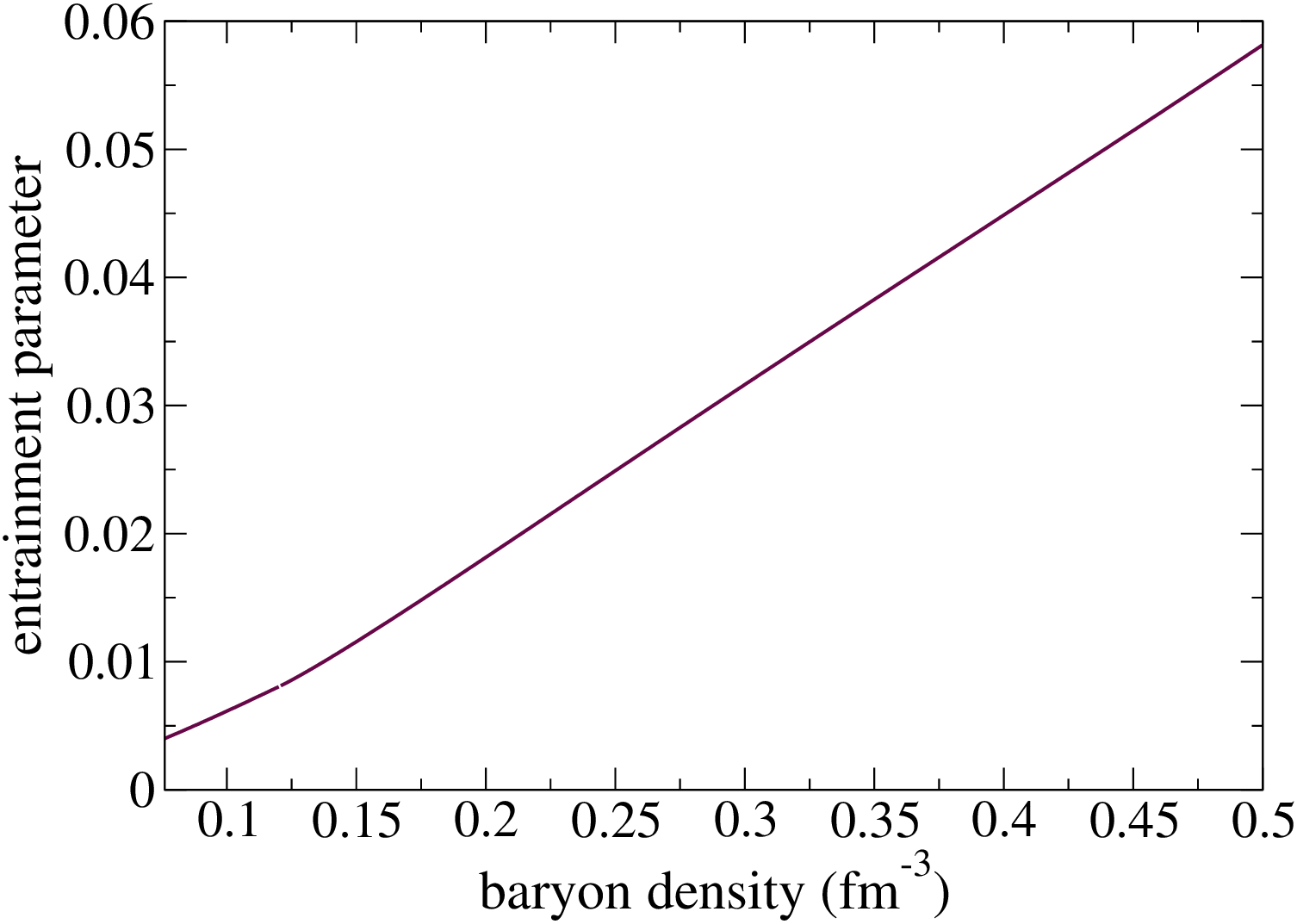}
\caption{\label{fig.smallepsilon}(Color online) Dimensionless entrainment parameter $\epsilon$
for $npe\mu$ matter in $\beta$ equilibrium (Skyrme SLy4 energy functional).}
\vskip 0.7cm
\end{figure}

\begin{figure}
\includegraphics[height=6cm,draft=false]{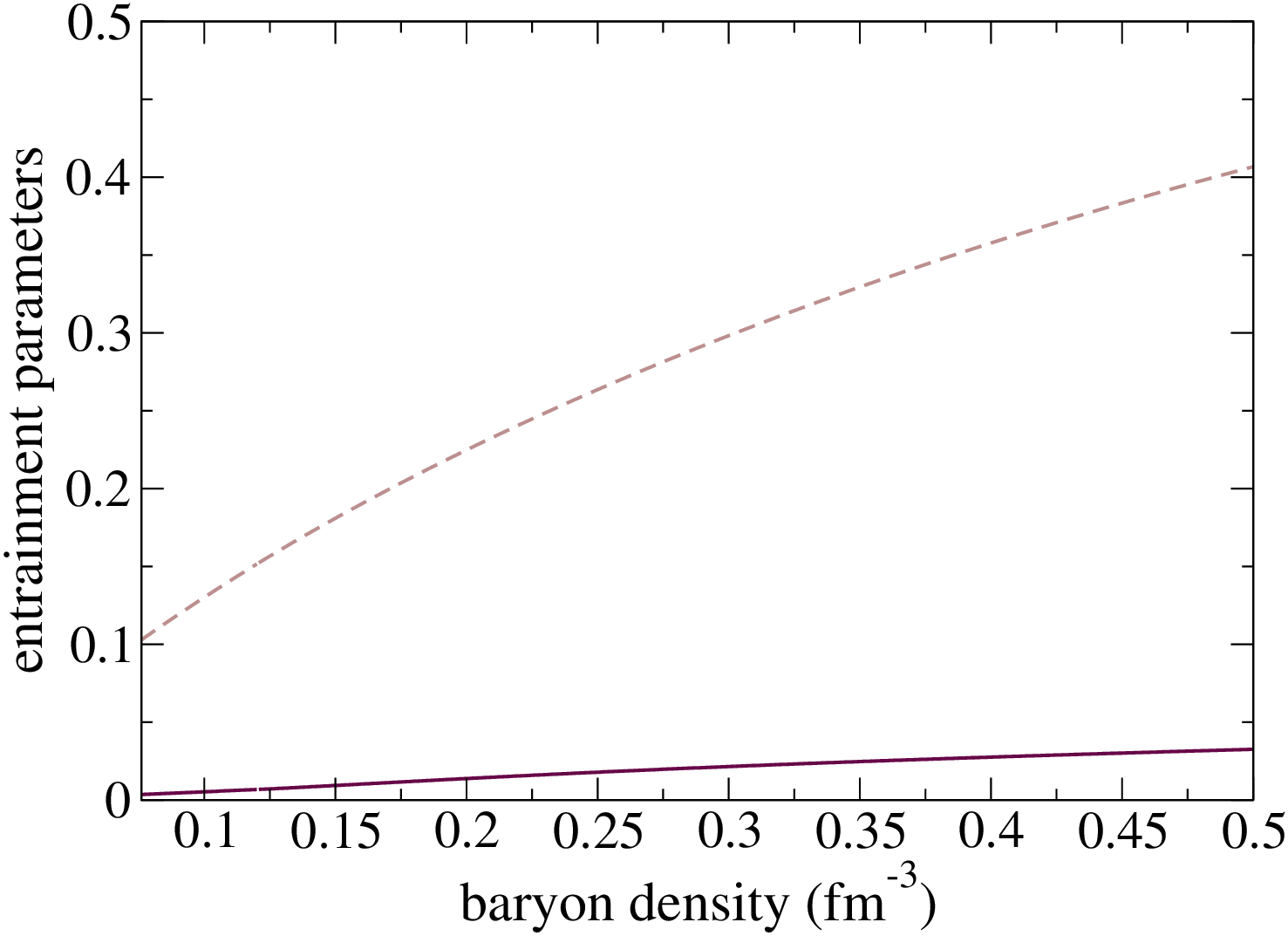}
\caption{\label{fig.epsilon} (Color online) Dimensionless  entrainment parameters
$\varepsilon_n$ (solid line) and $\varepsilon_p$ (dashed line) for
$npe\mu$ matter in $\beta$ equilibrium (Skyrme SLy4 energy
functional).}
\vskip 0.7cm
\end{figure}

\begin{figure}
\includegraphics[height=6cm,draft=false]{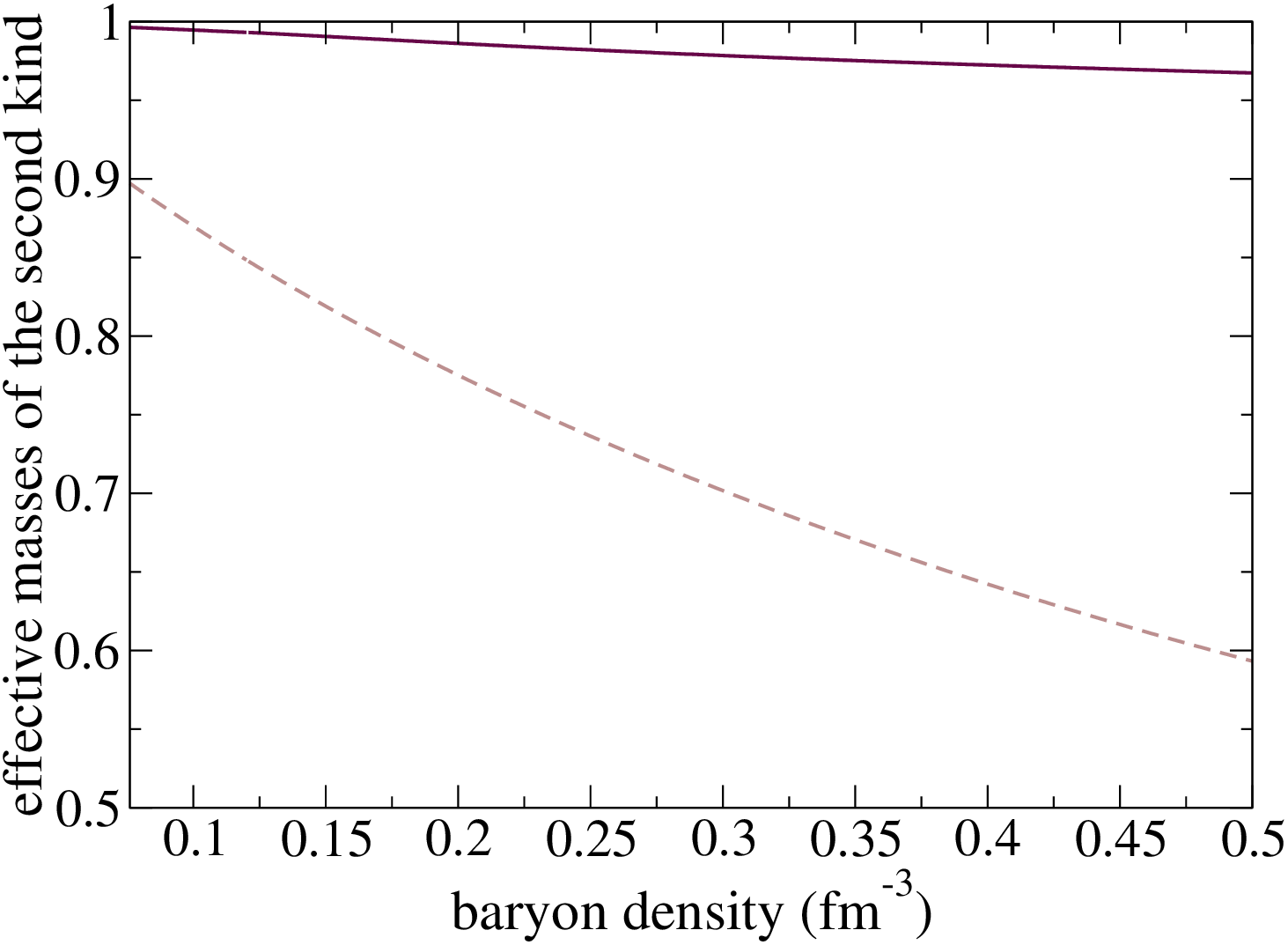}
\caption{\label{fig.effmass1}(Color online) Effective masses
$m_\star^n/m$ (solid line) and $m_\star^p/m$ (dashed line) for
$npe\mu$ matter in $\beta$ equilibrium (Skyrme SLy4 energy
functional).}
\vskip 0.7cm
\end{figure}

\begin{figure}
\includegraphics[height=6cm,draft=false]{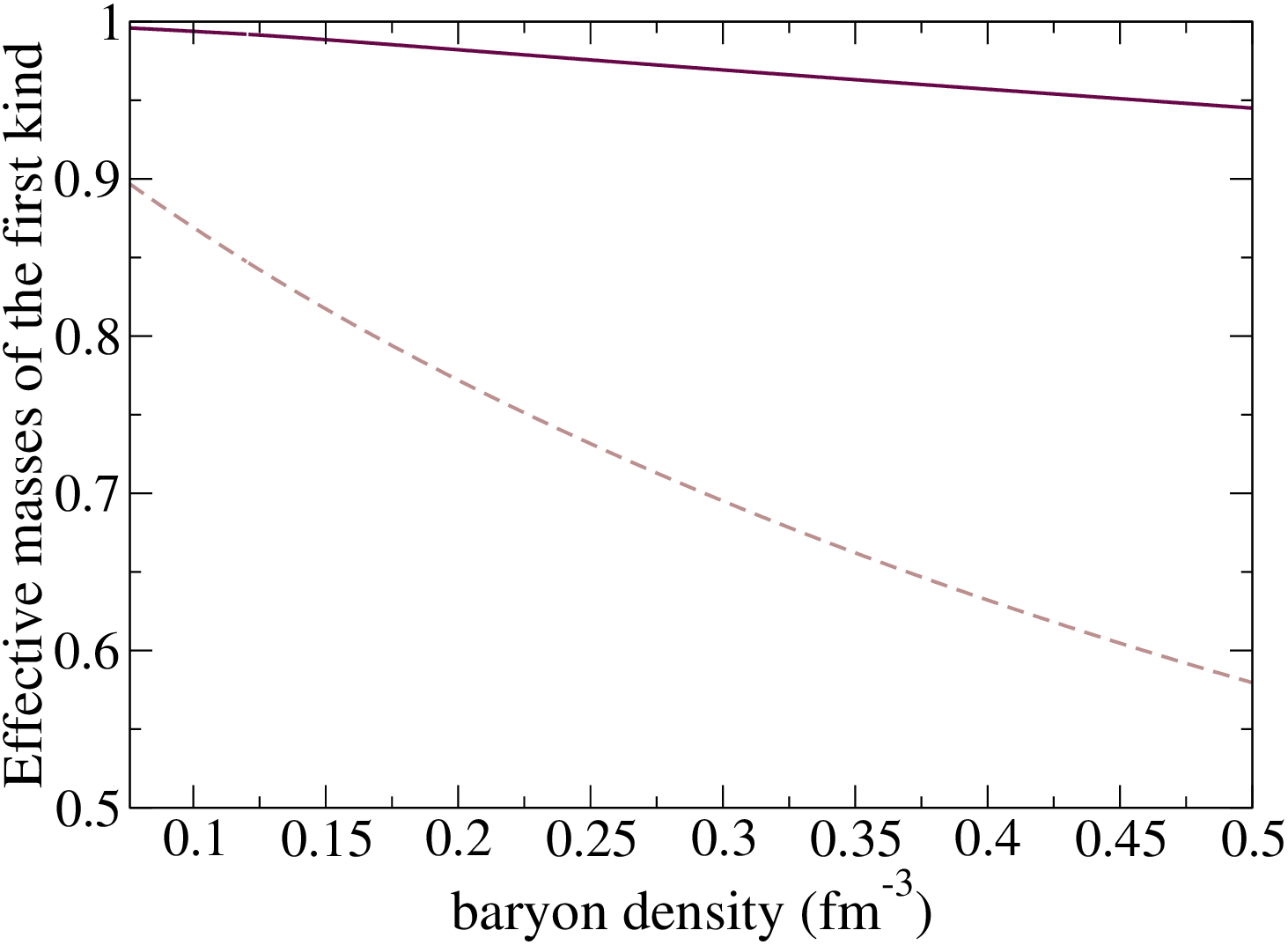}
\caption{\label{fig.effmass2} (Color online) Effective masses $m_\sharp^n/m$
(solid line) and $m_\sharp^p/m$ (dashed line) for $npe\mu$ matter
in $\beta$ equilibrium (Skyrme SLy4 energy functional).}
\vskip 0.7cm
\end{figure}

\begin{figure}
\includegraphics[height=6cm,draft=false]{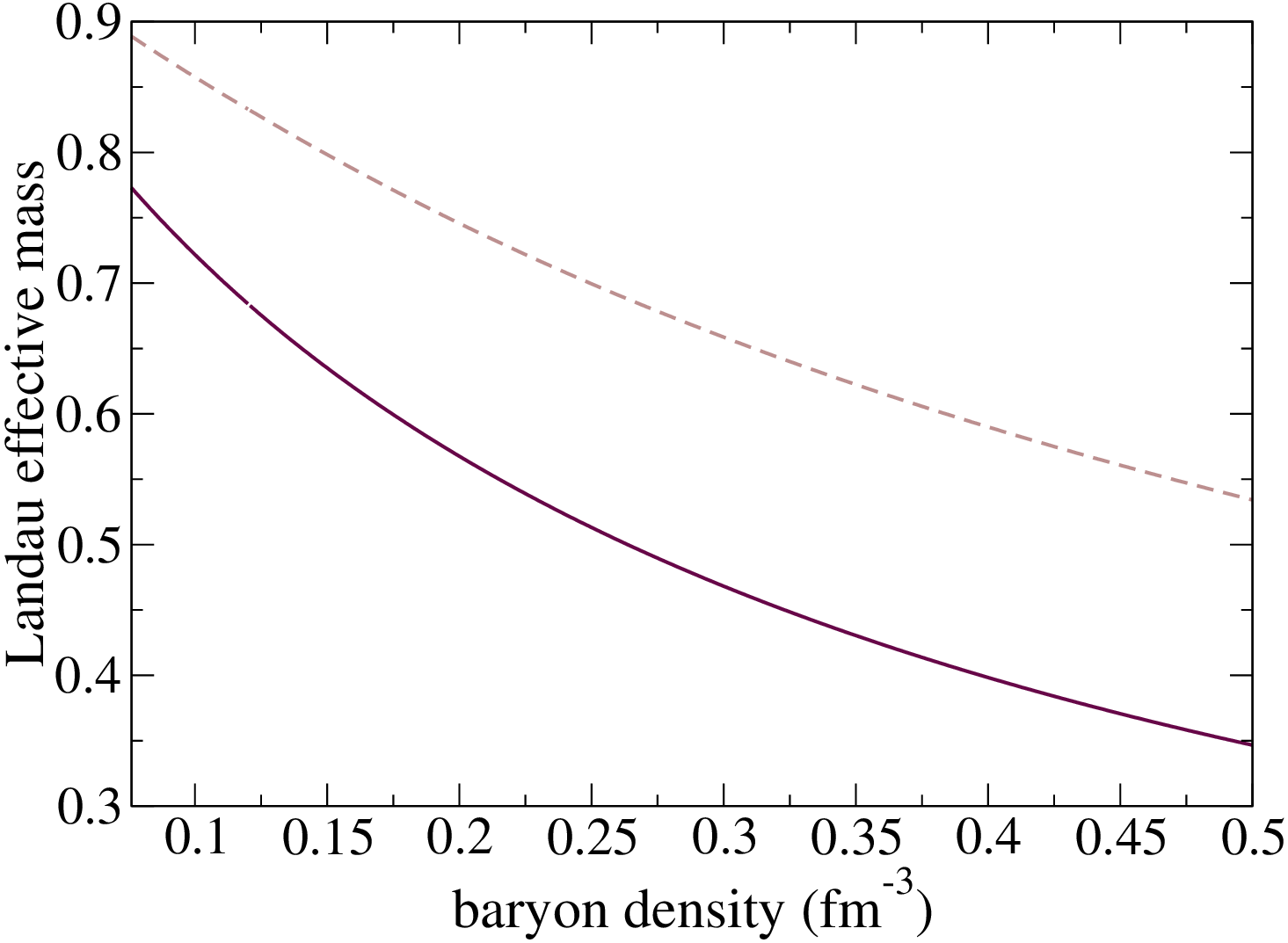}
\caption{\label{fig.landmass}(Color online) Landau effective masses $m_n^\oplus/m$
(solid line) and $m_p^\oplus/m$ (dashed line) for $npe\mu$ matter
in $\beta$ equilibrium (Skyrme SLy4 energy functional).}
\vskip 0.7cm
\end{figure}

The dimensionless entrainment parameters as defined by (\ref{eq.42}) and (\ref{eq.43}),
which have been widely used in neutron star simulations, are represented
on figures \ref{fig.smallepsilon} and \ref{fig.epsilon} respectively.
The $\star$- and $\sharp$- effective masses are shown on figures
\ref{fig.effmass1} and \ref{fig.effmass2} respectively. Due to the increase of the
proton fraction with the baryon density (see figure \ref{fig.composition}), the differences between the
two kinds of effective masses $m_\star^q$ and $m_\sharp^q$, which are negligible at the crust-core
boundary, become significant in deeper layers. We have also plotted on figure \ref{fig.landmass} the Landau
effective masses for comparison. As can be seen on those figures,
the various definitions of ``effective mass'' do not coincide. This concept should therefore
be carefully employed and the definition that has been adopted should always be
clearly specified.

\begin{figure}
\vskip 1cm
\includegraphics[height=6cm,draft=false]{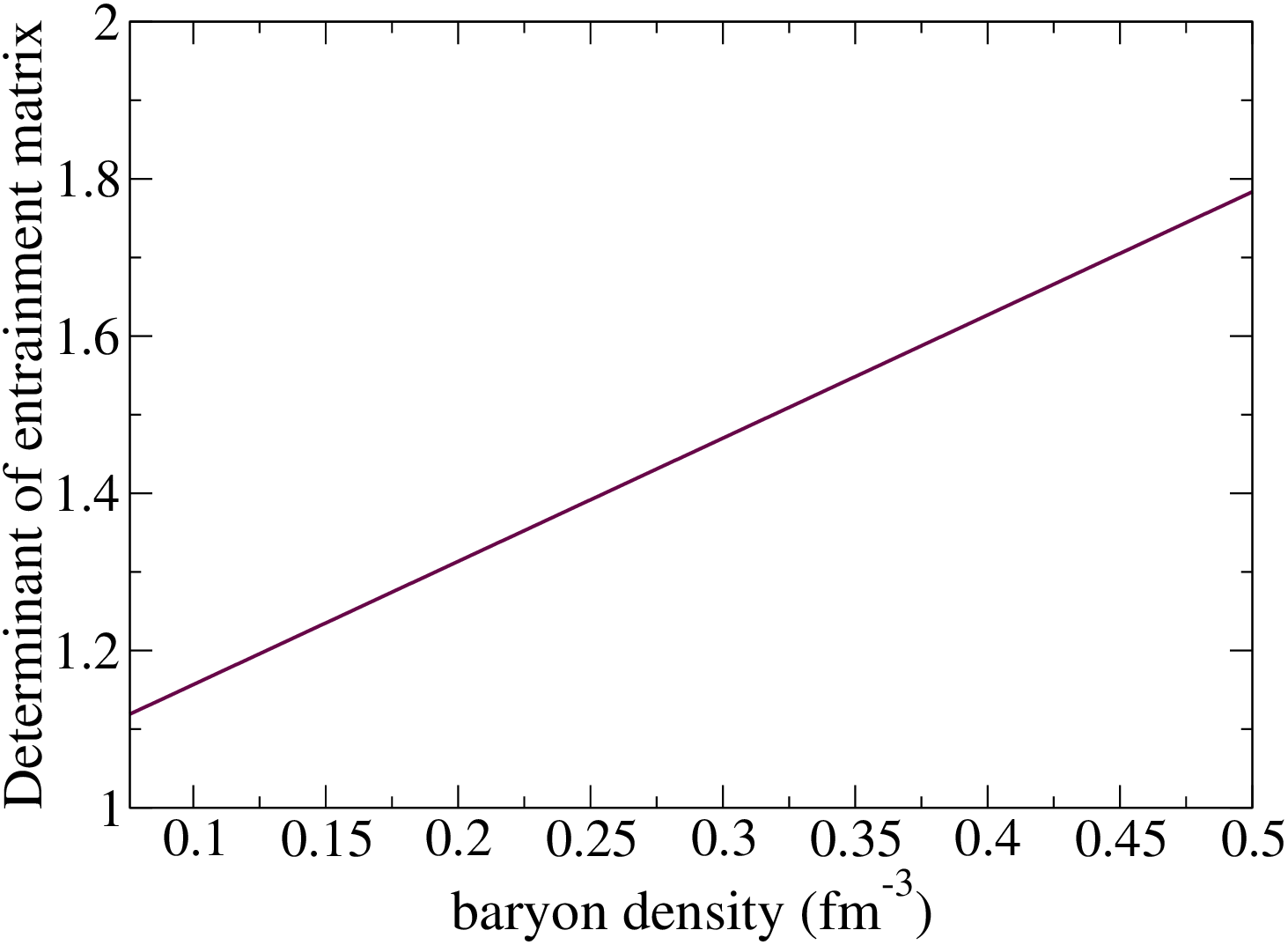}
\caption{\label{fig.gamma}(Color online) Dimensionless determinant $\Upsilon$ of the entrainment matrix
(Skyrme SLy4 energy functional).}
\end{figure}

We have finally shown on figure \ref{fig.gamma}, the dimensionless
determinant of the entrainment matrix
\begin{equation}
\Upsilon = \frac{\rho_{nn}\rho_{pp}-\rho_{np}^2}{\rho_n \rho_p}\, ,
\end{equation}
which appears in the perturbed hydrodynamical equations and which
is therefore important for the study of oscillation modes. In the present case,
this quantity is simply given by
\begin{equation}
\Upsilon = 1 - \widetilde{\alpha}_{np} \rho_{b} \, ,
\end{equation}
and is therefore quite remarkably independent of the nuclear asymmetry.

For comparison, we have computed the coefficient $\widetilde{\alpha}_{np}$ from which all the entrainment
parameters can be obtained, for the 27 Skyrme forces recommended by Rikovska Stone \textit{et al.}
\cite{rikovska03} for neutron star studies. The coefficient $\widetilde{\alpha}_{np}$ ranges
 from $0$ for the parametrisations SkT4 and SkT5 \cite{tondeur84} down to $-10.4168$ (in units m$_p^{-1}$.fm$^{-3}$)
for the parametrisation SV \cite{beiner75}. Let us also mention that the Skyrme SLy forces \cite{chabanat95, chabanat97, chabanat98, chabanat98err},
which have been widely employed in neutron star studies, yield coefficients around $\widetilde{\alpha}_{np}\sim -1.5$
except for the SLy230a force for which $\widetilde{\alpha}_{np}=-0.007360$.

Let us notice that these three forces SkT4, SkT5 and SLy230a are the only parametrisations for which the isovector effective mass, relevant
for the $T=1$ isovector electric dipole sum rule, was set equal to the bare nucleon mass (see Table VII of reference \cite{rikovska03}).
However, the isovector electric dipole ``giant resonance'' in nuclei consists essentially of relative motion of protons against neutrons, and the
sum rule constraint is therefore crucial for the entrainment effect in the infinite neutron-proton mixture.
The case $\widetilde{\alpha}_{np}=0$ implies that the effective masses $m_\sharp^q$ and $m_\star^q$
are equal to the bare one and therefore there is no entrainment. Reciprocally large negative values of
$\widetilde{\alpha}_{np}$ are associated with strong entrainment effects.

 In the present paper we considered only spin-unpolarized nuclear matter.
 We therefore did not discuss spin and spin-isospin instabilities that
plague many Skyrme forces at supranuclear densities \cite{margueron02, rios05}.
However, let us notice that for the SLy4 used
in our calculations, the ferromagnetic instability appears above the baryon density
$0.5~{\rm fm^{-3}}$, which is beyond the upper limit in our figures.


\section{Conclusion}

Analytical expressions for the entrainment matrix and  related
effective masses of a neutron-proton superfluid mixture at zero temperature
have been obtained within the non relativistic energy density functional theory.
In contrast to recent investigations within relativistic
mean field models \cite{comer03}, the entrainment
parameters have been found to be expressible by very simple
formulae which could be easily implemented in dynamical simulations of
neutron star cores. We have also clarified the link between various
definitions of effective masses that have been introduced in the literature. 

We have applied these formulae for Skyrme forces in order to evaluate
the entrainment matrix in the standard
model of the liquid core of neutron stars, composed of a mixture
of neutrons, protons, electrons and possibly muons in beta
equilibrium. Comparing the results with different Skyrme forces, we have found
that the entrainment parameters are quite sensitive to the adopted parametrisation.
The observations of the entrainment effects in neutron stars could therefore provide
new contraints on the construction of phenomenological nucleon-nucleon interactions
and shed light on the properties of strongly asymmetric nuclear matter.

\appendix
\section{Skyrme energy density functional coefficients}
The energy functional deduced from the Skyrme effective
interaction in the Hartree-Fock approximation has a similar form
as equation (\ref{eq.15}). The coefficients in the energy
functional (\ref{eq.15}) can thus be expressed in terms of the
parameters of the Skyrme interaction as follows. As a result of
local phase invariance   of the Skyrme forces
\cite{dobaczewski95}, the coefficients $C_T^j$ and $C_T^\tau$ are
related by
\begin{equation} C_T^j = - C_T^\tau \, .
\end{equation}
In terms of the parameters of the Skyrme interaction the
coefficients of the energy functional are given by \cite{bender03}
\begin{equation}
C_0^n \{ n_{\rm b}\} = \frac{3}{8}t_0+\frac{3}{48}t_3 n_{\rm
b}^\gamma
\end{equation}
\begin{equation}
C_1^n \{ n_{\rm b} \}=-\frac{1}{4}t_0(\frac{1}{2}+x_0) -
\frac{1}{24} t_3(\frac {1}{2}+x_3) n_{\rm b}^\gamma
\end{equation}
\begin{equation}
C_0^\tau = \frac{3}{16}t_1+\frac{1}{4} t_2( \frac{5}{4}+x_2)
\end{equation}
\begin{equation}
C_1^\tau =-\frac{1}{8} t_1(\frac{1}{2}+x_1)+\frac{1}{8}
t_2(\frac{1}{2}+x_2) \, .
\end{equation}

\begin{acknowledgments}
Nicolas Chamel acknowledges financial support from the  Lavoisier program of
the French Ministry of Foreign Affairs. This research has been
partially supported by the Polish MNiI Grant No. 1 P03D-008-27 and
the PAN/CNRS LEA Astro-PF.
\end{acknowledgments}

\bibliography{entrainment.bib}

\end{document}